\documentclass[11pt]{article}

\usepackage{amsmath, amssymb, amsthm}
\usepackage{algorithm}
\usepackage{algorithmic}
\usepackage{graphicx}
\usepackage{dsfont}
\usepackage{booktabs}
\usepackage{multirow}  % <--- ADD THIS LINE
\usepackage{hyperref}
\usepackage[numbers]{natbib}
\usepackage{geometry}
\usepackage{tikz}
\usetikzlibrary{arrows.meta, calc}
\usetikzlibrary{graphs, positioning, shapes.geometric}
\usetikzlibrary{decorations.markings}

\usepackage{optidef} % For optimization problem formatting
\usepackage{bbm} % For indicator function
\usepackage{natbib}

\title{Machine Learning-Guided Quota Optimization for Multi-Round Two-Sided Matching}

\author{Omid Khormali and Mia Pesavento\\
	Department of Mathematics\\
	University of Evansville}

\date{\today}

\begin{document}
	
	\maketitle

	\begin{abstract}
		This paper proposes an integrated framework for machine learning-guided quota optimization applied to multi-round sorority recruitment, a small two-sided
		market where approximately 100 potential new members (PNMs) are matched to three chapters through a structured process governed by the Release Figure
		Methodology (RFM). Our framework combines a Random Forest classifier trained on historical registration data to generate PNM--chapter compatibility scores,
		integer linear programs for Round~2 and Round~3 invitation quota optimization balancing fairness, coverage, and efficiency objectives, and a Deferred
		Acceptance algorithm for final matching. Applied to five years of de-identified recruitment data from a small Midwestern 
university, and working with only 282 matched training pairs, the compatibility model achieves a cross-validated ROC-AUC of 0.5822, reflecting the inherent difficulty of predicting social compatibility from pre-recruitment registration 
data in a data-limited setting. Because fairness and coverage constraints dominate quota allocation under noisy scores, the framework is designed to degrade gracefully when ML signal is weak. Optimized quotas closely align with actual coordinator decisions for active chapters, and the Deferred Acceptance algorithm replicates actual 2025 recruitment outcomes with a 96.4\% individual-level agreement rate and a 100\% match rate across 56 PNMs. An interactive web application implementing the framework is made available to recruitment coordinators. These results support the viability of data-driven approaches to small-market matching with broader applicability to other constrained two-sided markets.
	\end{abstract}


\section{Introduction}

Two-sided matching markets pervade critical areas of modern society, from medical residency placements to school choice programs. These markets pair agents from two distinct groups, such as students and universities, or workers and firms, based on mutual preferences. The design of effective matching mechanisms has profound implications for participant welfare, market stability, and institutional efficiency. While much of the theoretical and empirical literature focuses on large-scale markets with hundreds or thousands of participants, considerably less attention has been devoted to small-market matching problems, where the number of participants is limited and strategic behavior can have outsized effects on outcomes.\\

This paper examines matching in the context of sorority recruitment at small liberal arts colleges, a setting that exemplifies the unique challenges of small two-sided markets. Specifically, we study the formal recruitment process at a small private Midwestern university, where approximately 100 potential new members (PNMs) are matched to three sorority chapters over a structured three-round process. This market exhibits several features that distinguish it from well-studied large matching markets. The extreme size asymmetry (100 PNMs to 3 chapters), capacity constraints (each chapter can accept 35-40 members), and the prevalence of ``suicide bids'', cases where PNMs rank only a single chapter and accept the risk of remaining unmatched rather than joining a less--preferred organization.\\

The sorority recruitment process employs the Release Figure Methodology (RFM), a matching algorithm based on deferred acceptance principles \cite{gale1962college}. While RFM has proven successful in many contexts, small markets face distinctive challenges. First, with only three chapters, PNMs have severely limited options, and a single misranking can result in no match. Second, strategic behavior becomes more consequential when each decision affects a larger fraction of the market. Third, the prevalence of suicide bids—where PNMs express only a single preference, creates inefficiencies, as these PNMs face high risks of remaining unmatched if their preferred chapter does not select them.

\paragraph{Research Questions}

This study investigates the application of machine learning and optimization techniques to improve matching outcomes in small-market sorority recruitment. Our analysis addresses three interrelated questions. First, we examine whether machine learning models can accurately predict compatibility between PNMs and chapters using only registration data collected before the recruitment process begins. We train binary classifiers on historical recruitment outcomes to generate compatibility scores for each PNM-chapter pair, providing a data-driven foundation for optimizing invitation decisions across recruitment rounds. Understanding the predictive accuracy of these models is essential for designing effective quota allocation strategies.\\

Second, we develop an optimization framework for determining invitation quotas that chapters should issue in Rounds 2 and 3 of the recruitment process. Using ML-predicted compatibility scores and observed reciprocal interest rates, we formulate integer linear programs that balance fairness constraints, coverage requirements, and efficiency objectives. This framework addresses a key operational challenge faced by recruitment coordinators: how to allocate limited invitation slots across chapters to maximize match quality while maintaining equity.\\

Third, we evaluate the final matching outcomes produced by the Deferred Acceptance algorithm under our optimized invitation quotas. We analyze match quality metrics including the number of matched PNMs, the distribution of suicide bids, preference satisfaction rates, and adherence to fairness constraints on chapter capacities. By comparing outcomes across different recruitment years and examining the role of ML-guided quota optimization, we assess whether data-driven approaches can improve upon traditional ad-hoc methods for managing small-market recruitment processes.

\paragraph{Paper Organization}

The remainder of this paper proceeds as follows. Section 2 provides background on the sorority system in American higher education and describes the multi-round recruitment process in detail, including the roles of invitation quotas, preference revelation, and final matching. Section 3 develops our methodological framework, presenting the machine learning compatibility prediction model, the optimization problems for Rounds 2 and 3 invitation quotas, and the Deferred Acceptance algorithm used for final matching. Section 4 presents our empirical results, analyzing match outcomes, and the effectiveness of ML-guided quota optimization. Section 5 concludes with a discussion of implications for recruitment practice and directions for future research.

\section{Background}

%\subsection{The Sorority System in American Higher Education}

The sorority system emerged in the mid-nineteenth century as women began entering American higher education in larger numbers. As women entered universities where they remained a small minority, many sought to establish organizations parallel to the fraternities and secret societies that male students had formed. The first women's secret societies were founded in the 1850s, and by the late nineteenth century, sixteen women's fraternities\footnote{Historically, women's Greek organizations were called ``fraternities.'' The term ``sorority'' was coined later but is now more commonly used.} had formed across the country \cite{sorority_history}.\\

Today, the National Panhellenic Conference (NPC) serves as the umbrella organization for 26 inter/national women's sororities \cite{npc_about}. Member organizations emphasize scholarship, leadership, service, and sisterhood. Formal recruitment, the structured process through which women join sororities, occurs annually at universities across the United States, with processes varying by institution size and local Panhellenic community structure. At larger universities, recruitment may involve ten to fifteen or more chapters, while smaller institutions may host as few as three or four active chapters.\\

%\subsection{Study Setting and Institutional Context}

Our study examines recruitment data from a small private liberal arts university in the Midwestern United States with an undergraduate enrollment of approximately 2,000 students. The institution is home to three National Panhellenic Conference (NPC) member sororities that participate in formal recruitment at the beginning of each fall semester. For confidentiality, we refer to these chapters as Chapter A, Chapter B, and Chapter C throughout this paper.\\

The small scale of this Greek system, three chapters serving a modest-sized student body, creates a matching market that differs substantially from larger universities where eight to fifteen or more sororities may participate in recruitment. This constrained choice set has important implications for participant strategy, match quality, and the prevalence of unmatched outcomes. The limited number of options means that PNMs face a fundamentally different decision problem than those at larger institutions, and strategic choices such as suicide bids carry greater risk.

\subsection{The Formal Recruitment Process}

Formal sorority recruitment at the University of Evansville is a structured, multi-round process governed by the Release Figure Methodology (RFM) \cite{rfm_update}, an algorithmic matching system based on principles of deferred acceptance. The process unfolds over one week and consists of three rounds followed by a final matching and bid distribution. We describe each component in detail below.

\subsubsection{Round 1: Introduction Round}

The recruitment process begins with Introduction Round, which spans two evenings. During this round, every registered PNM visits all three sorority chapters. Conversations focus on getting acquainted: hometowns, academic interests, extracurricular activities, and personal backgrounds. Each chapter hosts multiple short parties throughout the evening to accommodate all PNMs.\\

Following Round 1, each chapter receives a quota from the RFM specialist specifying how many PNMs they may invite to Round 2. Chapters then rank PNMs and create their invitation lists. Importantly, at this stage, PNM preferences are \emph{not} considered—invitation decisions are unilateral. Each PNM receives a schedule indicating which chapters have invited her to return for Round 2.

\subsubsection{Round 2: Philanthropy and Sisterhood Round}

Round 2 takes place on a single evening. PNMs visit only those chapters that invited them back, with each PNM potentially visiting anywhere from zero to three chapters. The modal outcome is visiting all three chapters, though some PNMs receive fewer invitations.\\

Conversations during Round 2 center on each chapter's philanthropic activities, community service initiatives, and the nature of sisterhood within the organization. Chapters are also required to disclose membership costs during this round, providing PNMs with information about the financial commitment required.\\

After Round 2, mutual preference information enters the matching process for the first time. The RFM specialist provides each chapter with a quota for Round 3 invitations. PNMs simultaneously submit rankings indicating which chapters they wish to visit for Round 3. In the current three-chapter system, PNMs assign a rank of ``1'' to their top two choices and ``2'' to their least-preferred option (or the chapter they do not wish to visit). The RFM algorithm then performs a matching to determine each PNM's Round 3 schedule, taking into account both chapter invitation lists and PNM preferences.

\subsubsection{Round 3: Preference Night}

Round 3, known as Preference Night, represents the culmination of the recruitment process. PNMs visit at most two chapters during this final round, with some visiting only one if they received just one invitation or chose to rank only one chapter highly enough. Conversations become more intimate and substantive, often focusing on values alignment and the decision to join.\\

Following Round 3, chapters create their \emph{bid lists}—complete rankings of all PNMs who visited them during Preference Night. These rankings represent each chapter's true preferences over potential members, ordered from most to least desired. Simultaneously, PNMs complete the Membership Recruitment Acceptance Binding Agreement (MRABA), a legally binding document in which they list their chapter preferences in order. Critically, PNMs may list one or two chapters; listing only one chapter constitutes a \emph{suicide bid} or \emph{single preference}.\\

The MRABA is binding: a PNM who lists a chapter on her MRABA and receives a bid from that chapter \emph{must} accept it or become ineligible to join any NPC sorority until the next recruitment cycle (one year later). This mechanism is designed to ensure commitment but creates strategic tension for PNMs, particularly those who feel strongly drawn to a single chapter.

\subsubsection{Final Matching and Bid Day}

Using the chapters' complete bid lists and PNMs' MRABA preferences, the RFM specialist executes a final matching algorithm. The algorithm is a variant of the deferred acceptance mechanism, designed to produce stable matchings while respecting capacity constraints and the binding nature of PNM preferences \cite{rfm_update}.\\

On Bid Day, chapters receive their new member lists, and PNMs simultaneously open bid cards revealing which chapter, if any, has extended them an invitation. Matched PNMs then join their new chapters for celebration events. Unmatched PNMs—those who either submitted suicide bids for chapters that did not select them, or whose preferences could not be satisfied given capacity constraints—receive no bid and must wait until the next recruitment cycle or pursue other affiliation opportunities.

\subsection{Key Terminology and Concepts}

For clarity, we define several terms central to this study. A \textit{Potential New Member (PNM)} is a woman registered for and participating in formal sorority recruitment who is not currently affiliated with an NPC member organization. A \textit{chapter} refers to the local, campus-based unit of a national or international sorority; in this study, we focus on the three NPC chapters at the participating institution. A \textit{bid} is a formal invitation to join a sorority, extended on Bid Day as the outcome of the matching process.\\

A \textit{suicide bid}, or single preference, refers to the strategic choice by a PNM to list only one chapter on her MRABA preference form despite having the option to list two. This reflects strong preference intensity but carries the risk of remaining unmatched if the preferred 
chapter does not extend a bid. The \textit{Release Figure Methodology (RFM)} is the algorithmic matching system used by the National Panhellenic Conference to manage sorority recruitment; it is based on deferred acceptance principles and is designed to produce stable matchings while 
managing capacity constraints across multiple rounds~\cite{rfm_update}. \\

A \textit{matching} is an assignment of PNMs to chapters such that capacity constraints are satisfied, and is said to be stable if no PNM-chapter pair exists who would both prefer each other over their current assignments~\cite{gale1962college}. Finally, \textit{capacity} refers to the target number of new members a chapter aims to recruit, typically ranging from 35 to 40 in our setting.\\

The analysis of two-sided matching markets has a rich intellectual history spanning economics, computer science, and operations research. We review three streams of literature relevant to our study: classical matching theory, applications of machine learning to matching problems, and empirical work on small market dynamics.

\section{Literature Review}

The modern theory of two-sided matching markets originates with Gale and 
	Shapley~\cite{gale1962college}, who introduced the concept of a stable matching and proved that one always exists in any finite two-sided market. Their Deferred Acceptance (DA) algorithm guarantees a stable outcome that is optimal for the proposing side, and this result earned Shapley and Roth the 2012 Nobel Memorial Prize in Economic Sciences~\cite{nobelprize2012}. Building on this foundation, 
	Roth~\cite{roth1982economics} established that no stable mechanism can be strategy-proof for both sides simultaneously, and that truthful preference revelation is a weakly dominant strategy for the proposing side under DA. The empirical importance of stability was documented in Roth~\cite{roth1984evolution}, who showed that markets organized to produce stable outcomes survive and function well over time, while those employing unstable algorithms tend to unravel. Roth and Sotomayor~\cite{roth1990twosided} provide the most comprehensive theoretical treatment of the field, including the Rural Hospitals Theorem and extensions to many-to-one markets where one side, such as colleges or sorority chapters, accommodates multiple agents subject to capacity constraints.\\
	
	These theoretical foundations have been applied to an expanding range of 
	real-world clearinghouses. Roth and Peranson~\cite{roth1999redesign} redesigned the National Resident Matching Program (NRMP), the centralized clearinghouse that annually places approximately 20,000 medical graduates into residency positions, demonstrating that in large markets strategic manipulation is rare and stable matchings are reliably achievable. Abdulkadiro\u{g}lu and S\"{o}nmez~\cite{abdulkadiroglu2003school} extended the framework to public school choice, showing that the prevalent Boston mechanism gave students incentives to misrepresent their preferences, and proposing the student-optimal DA mechanism as a strategy-proof alternative subsequently adopted by New York City and Boston. Both applications underscore a point directly relevant to our setting: 
	mechanism choice has profound consequences for participant incentives and match quality, and these consequences are amplified in smaller markets where each agent's decision affects a larger fraction of the total pool.\\
	
	The specific context of sorority recruitment was analyzed by Mongell and 
	Roth~\cite{mongell1991sorority}, who examined the Preferential Bidding System (PBS) used historically in sorority recruitment. Unlike DA, the PBS does not guarantee stable outcomes. Analyzing data from twenty-one recruitment cycles, the authors found that the system rarely failed in practice, attributing this to the widespread tendency of rushees to submit single-preference bids, which effectively removed problematic preference configurations. This finding is directly relevant to our study, where suicide bids remain prevalent and constitute a central operational challenge. Their work also establishes the closest empirical precedent to ours; no subsequent study has applied data-driven methods to the operational challenges of small-market sorority recruitment.\\
	
	A growing literature has explored machine learning and optimization techniques in matching contexts. Hitsch et al.~\cite{hitsch2010matching} showed that revealed preferences in online dating markets can be inferred from behavioral data and used to improve match recommendations, while Hsieh~\cite{hsieh2019prediction} found that standard ML recommenders create congestion in two-sided platforms by concentrating recommendations on popular options, motivating quota-aware allocation approaches. On the optimization side, Bir\'{o} et al.~\cite{biro2016integer} developed integer programming formulations for college admissions problems with ties and common quotas, and Roth, Rothblum, and Vande Vate~\cite{roth1993stable} established that the set of stable matchings 
	can be characterized as the integer points of a linear programming polytope. Taken together, these contributions demonstrate that data-driven quota optimization is a principled and practically viable complement to stable matching algorithms, particularly in constrained, small-market settings where ad hoc coordinator decisions are difficult to benchmark and improve upon systematically. Our work brings these three streams together in the context of multi-round sorority recruitment.

\section{Method}

    This study uses de-identified administrative recruitment data from five years (2021-2025) provided by a small Midwestern university. All personally identifiable information has been removed, and the institution has been anonymized at the request of university administrators. The research was classified as exempt from IRB review as it involves analysis of existing de-identified administrative records.\\

    Our approach develops and implements a comprehensive solution to the challenge 
faced by recruitment coordinators in determining optimal invitation quotas across multiple rounds while balancing fairness, coverage, and match quality. We develop a four-step framework that combines machine learning and optimization to align with the sorority recruitment process. The four steps are (1) ML-based compatibility prediction from registration data, (2) Round~2 
invitation quota optimization, (3) Round~3 invitation quota optimization, and (4) final matching with capacity determination. 
	
	\subsection{Notation and Definitions}
	
	Let $\mathcal{P} = \{p_1, p_2, \ldots, p_n\}$ denote the set of $n$ Potential New Members (PNMs) and $\mathcal{C} = \{c_1, c_2, \ldots, c_m\}$ denote the set of $m$ chapters participating in recruitment. The recruitment process consists of multiple rounds as follows.
	
	\begin{itemize}
		\item [--] Registration Phase: PNMs register and provide their information.
		\item [--] Round 1: All PNMs meet with all $m$ chapters over multiple nights for initial introductory conversations without sending any invitations.
		\item [--] Round 2: Based on Round 1 impressions, chapters send selective invitations to PNMs, and invited PNMs visit chapters and then rank those they visited.
		\item [--] Round 3: Chapters send narrower invitations which are subset of Round 2, and PNMs visit invited chapters. Both PNMs and chapters rank each other, and PNMs rank up to 2 chapters.
		\item [--] Final Matching: A third-party coordinator (for convenience we call the coordinator Paula) runs a matching algorithm using PNM and chapter preference lists to generate final matches.
	\end{itemize}
	
	Throughout the process, Paula makes several {\it key decisions} which are as follows.
	\begin{enumerate}
		\item Determines invitation quotas for each chapter in Round 2
		\item Determines invitation quotas for each chapter in Round 3 
		\item Specifies how many PNMs each chapter should rank for final matching
		\item Runs the final matching algorithm
		\item Reports final chapter capacities
	\end{enumerate}
	\vspace{.09in}
	Our optimization framework focuses on steps 1 and 2 that determine optimal invitation quotas for Rounds 2 and 3. For step 3, we assume each chapter ranks all PNMs who were invited in Round 3, as this simplification has minimal impact on match outcomes while avoiding substantial additional computational complexity. \\
	
	For rounds $r \in \{2, 3\}$, let $x_i^{(r)}$ denote the number of invitations chapter $c_i$ sends in round $r$. Our goal is to determine optimal values for these invitation quotas to maximize match quality while ensuring fairness and adequate PNM coverage.\\
	
	The key constraints on invitation quotas are as follows.\\
	
	\underline{Coverage Constraints.} In Round 2, PNMs may receive invitations from 1, 2, or 3 chapters, with most receiving 2-3 invitations. In Round 3, PNMs receive 1-2 invitations, with most receiving 2. To ensure adequate coverage, we set the following conditions.
	\begin{align*}
		2n &\leq \sum_{i=1}^{m} x_i^{(2)} \leq 3n \\
		n &\leq \sum_{i=1}^{m} x_i^{(3)} \leq 2n 
	\end{align*}
	
	These bounds ensure that each PNM typically receives about 2–3 invitations in Round 2 and 1–2 invitations in Round 3, which matches the usual recruitment pattern.\\
	
	\underline{Fairness Constraints.} Historical recruitment data shows that chapter invitation quotas in Rounds 2 and 3, as well as chapter capacities, are generally well balanced across chapters, with only minor differences. Let $\bar{x}^{(r)} = \frac{1}{m}\sum_{i=1}^{m} x_i^{(r)}$ denote the average invitation quota in round $r$. To capture the chapter capacities imbalance observed in past recruitment cycles, we define the fairness parameter $\rho$ as the {\it maximum proportional deviation} of chapter capacities from their mean.
	\[
	\rho = \max_{1 \le i \le m}
	\left|\frac{C_i - \bar{C}}{\bar{C}}\right|.
	\]
	
	Using this value ensures that no chapter’s quota differs from the average by more than the largest deviation we have historically observed. Based on this, we consider the fairness constraint as follows.
	\[
	(1 - \rho)\bar{x}^{(r)} \le x_i^{(r)} \le (1 + \rho)\bar{x}^{(r)},
	\quad \forall i \in \{1,\ldots,m\}.
	\]
	
	Note that for convenience, instead of measuring fairness based on historical variation in invitation quotas for round $r$, we estimate it using historical variation in chapter capacities, from which we compute the proportional deviation parameter $\rho$. \\
	
	\underline{Sequential Constraint.} Round 3 invitation lists are smaller than Round 2 and typically represent a subset of Round 2 invitees. Therefore, we set the following condition.
	\begin{equation}
		x_i^{(3)} \leq x_i^{(2)}, \quad \forall i \in \{1, \ldots, m\} \tag*{}
	\end{equation}
	
	\underline{Integrality.} All invitation quotas must be integers.
	\begin{equation}
		x_i^{(r)} \in \mathbb{Z}_+, \quad \forall i, r \tag*{}
	\end{equation}
	
	In practice, Paula asks each chapter to provide not only a primary invitation list of
	size $x_i^{(r)}$, but also two supplementary lists of fixed size $k$.
	\begin{itemize}
		\item[--] Flex+ List: $k$ additional PNMs the chapter would like to invite (backup candidates),
		\item[--] Flex-- List: $k$ PNMs from the primary list the chapter would consider removing if necessary.
	\end{itemize}
	
	These flex lists give Paula limited flexibility to make minor adjustments to balance
	the system. In the data provided to us, the value is $k = 8$, and we use this value in our implementation, although any other fixed size could be selected depending on policy or future recruitment practices. Our optimization model, however, focuses on determining the primary invitation quota $x_i^{(r)}$ for each chapter.

	\subsection{Step 1: ML-Based Compatibility Prediction}
	
	% During the registration phase, PNMs provide their information which including:
	
	% \textcolor{red}{complete the list}
	% \begin{itemize}
	% 	\item Birth date (Year)
	% 	\item Campus Address (Residence Hall) 
	% 	\item Year in College
	% 	\item Major
 %        \item College GPA
	% 	\item High School Name
	% 	\item High School GPA
	% 	\item Activities Involved In
 %        \item Dietary Restrictions
	% 	\item T-Shirt Size
	% \end{itemize}
    During the registration phase, PNMs provide demographic and academic information including birth year, campus residence hall, year in college, major, college GPA (if applicable), high school name, high school GPA, extracurricular activities, dietary restrictions, and t-shirt size. We construct a feature vector $\mathbf{f}_p \in \mathbb{R}^d$ for each PNM $p$ from this registration data. \\
 %    Similarly, each chapter $c$ is characterized by a profile vector $\mathbf{h}_c \in \mathbb{R}^k$ representing the chapter's organizational characteristics which including:
	
	% 	\textcolor{red}{complete the list}
	% \begin{itemize}
	% 	\item Where Active Members Went to High School
	% 	\item Active Members Majors
	% 	\item Average Chapter GPA
	% 	\item Cost of Membership
 %        \item Activities Active Members are Involved In
	% \end{itemize}
	% \ \\

    Similarly, each chapter $c$ is characterized by a profile vector $\mathbf{h}_c \in \mathbb{R}^k$ constructed from chapter-level data including average member GPA, annual membership cost, chapter size, and aggregated member characteristics (high school origins, academic majors, extracurricular activities). Chapter features are constructed by aggregating information across current active members: categorical features (high schools, majors) are represented as distribution vectors, continuous features (GPA, cost, size) are normalized, and text features (activities) are aggregated and encoded using the same method as PNM activity features.\\

	For each PNM-chapter pair $(p, c)$, we create a combined feature vector as follows.
	\begin{equation}
		\mathbf{z}_{pc} = [\mathbf{f}_p; \mathbf{h}_c] \in \mathbb{R}^{d+k} \tag*{}
	\end{equation}
	where $[\cdot;\cdot]$ denotes concatenation of PNM features and chapter features.\\

	Using historical recruitment data $\mathcal{D} = \{(\mathbf{z}_{pc}, y_{pc})\}$ where $y_{pc} \in \{0,1\}$ indicates whether PNM $p$ was successfully matched with chapter $c$ in past recruitment cycles, we train a binary classifier $\hat{y}_{pc} = f_\theta(\mathbf{z}_{pc})$. We evaluate multiple classification models including 
	Logistic Regression, Random Forest Classifier, and Gradient Boosting (XGBoost). We choose the model using cross-validated AUC-ROC scores and check calibration to make sure the predicted probabilities are realistic.\\
	
	For the current recruitment cohort, we construct a compatibility matrix $\mathbf{S} \in \mathbb{R}^{n \times m}$, where each entry $S_{pc} \in [0,1]$ represents the predicted probability that PNM $p$ is compatible with chapter $c$, and we have
	\[
	S_{pc} = \mathbb{P}(y_{pc} = 1 \mid \mathbf{z}_{pc})  \in [0,1].
	\]

	This matrix captures the predicted compatibility between each PNM and each chapter using only registration data. It serves as input to the optimization problems in Steps 2 and 3.\\

    The training dataset contains one row per PNM--chapter pair, yielding $n \times m$ rows per recruitment year, where $n$ is the number of PNMs and $m$ is the number of chapters. Since each PNM is matched to exactly one chapter, the ratio of negative to positive instances is structurally fixed at $m-1$ to $1$, producing a 3:1 class imbalance with four chapters. 
This imbalance is an inherent property of the matching structure rather than a data collection artifact, and standard resampling remedies are 
therefore not applicable. We retain predicted probabilities as relative compatibility signals rather than binary predictions, which is appropriate given this structural constraint.
	
	\subsection{Step 2: Round 2 Invitation Quota Optimization}
	
	After Round 1 where all PNMs meet all chapters, Paula must determine how many PNMs each chapter should invite to Round 2. This decision directly impacts PNM coverage.\\
	
	For each chapter $c_i$, we calculate a popularity weight $w_i^{(2)}$ based on ML compatibility scores. We define popularity as the proportion of PNMs for whom chapter $c_i$ has the highest predicted compatibility:
	\begin{equation}
		w_i^{(2)} = \frac{1}{n}\sum_{p=1}^{n} \mathbbm{1}\left[\arg\max_{j \in \{1,\ldots,m\}} S_{pj} = i\right]
	\end{equation}
	
	This weight is the proportion of PNMs for whom chapter $c_i$ is the top ML-predicted match. Chapters with higher $w_i^{(2)}$ are expected to see more PNMs return in Round 2, so they may need fewer initial invitations.
	
	\subsubsection{Optimization Models}
	
	We formulate an optimization model for Round 2 invitation quotas. The model balances different objectives while respecting coverage and fairness constraints.
	
	\paragraph{Model: Minimize Weighted Sum (Efficiency-Focused)}
	\begin{mini!}|l|
		{x_1^{(2)}, \ldots, x_m^{(2)}}
		{\sum_{i=1}^{m} w_i^{(2)} x_i^{(2)}}{}{}\tag*{}
		\addConstraint{2n}{\le \sum_{i=1}^{m} x_i^{(2)} \le 3n}{\nonumber}
		\addConstraint{(1-\rho)\bar{x}^{(2)}}{\le x_i^{(2)} \le (1+\rho)\bar{x}^{(2)}}{,\quad \forall i \nonumber}
		\addConstraint{x_i^{(2)}}{\in \mathbb{Z}^+}{,\quad \forall i \nonumber}
	\end{mini!}

	This model helps chapters with high $w_i^{(2)}$ get fewer invitations because they attract more PNMs back. Chapters with lower $w_i^{(2)}$ receive more invitations to make up for weaker return rates.
	
	% \paragraph{Model B: Maximize Coverage (Coverage-Focused)}
	% \begin{maxi!}|l|
	% 	{x_1^{(2)}, \ldots, x_m^{(2)}}
	% 	{\sum_{i=1}^{m} x_i^{(2)}}{}{}\tag*{}
	% 	\addConstraint{\sum_{i=1}^{m} x_i^{(2)}}{\leq 3n}{\nonumber}
	% 	\addConstraint{(1-\rho)\bar{x}^{(2)}}{\leq x_i^{(2)} \leq (1+\rho)\bar{x}^{(2)}}{,\quad \forall i \nonumber}
	% 	\addConstraint{x_i^{(2)}}{\in \mathbb{Z}^+}{,\quad \forall i \nonumber}
	% \end{maxi!}

	% This model maximizes the total number of invitations, within fairness and capacity limits, to increase the chance that all PNMs receive multiple invitations.\\
	
	The model is an integer linear program that can be solved efficiently using standard optimization tools such as PuLP, a Python library for linear programming.
	
	\subsection{Step 3: Round 3 Invitation Quota Optimization}
	
	After Round 2, PNMs submit rankings of the chapters they visited. Paula then determines Round 3 invitation quotas, which are smaller and more selective than Round 2. We observe which PNMs ranked each chapter. For each chapter $c_i$, we calculate the empirical reciprocal rate as follows.
	\begin{equation}
		r_i = \frac{v_i}{x_i^{(2)}}
	\end{equation}
	where $v_i$ is the number of PNMs who included chapter $c_i$ in their Round 2 ranking regardless of ranking position, and $x_i^{(2)}$ is the number of invitations chapter $c_i$ sent.	For Round 3 optimization, we define weights based on reciprocal rates
	\begin{equation}
		w_i^{(3)} = r_i.
	\end{equation}
	
	Chapters with high reciprocal rates attract many PNMs back, so Model A gives them fewer invitations. Chapters with lower reciprocal rates receive more invitations to ensure enough PNMs return.
	
	\subsubsection{Optimization Models for Round 3}
	
	The optimization model for Round 3 mirrors the structure of the Round 2 model, with updated constraints reflecting the narrowing of the recruitment funnel.

	\paragraph{Model: Minimize Weighted Sum}
	\begin{mini!}|l|
		{x_1^{(3)}, \ldots, x_m^{(3)}}{\sum_{i=1}^{m} w_i^{(3)} x_i^{(3)}}{}{}\tag*{}
		\addConstraint{n}{\leq \sum_{i=1}^{m} x_i^{(3)} \leq 2n}{}{\nonumber}
		\addConstraint{(1-\rho)\bar{x}^{(3)}}{\leq x_i^{(3)} \leq (1+\rho)\bar{x}^{(3)}}{, \quad \forall i}{\nonumber}
		\addConstraint{x_i^{(3)}}{\leq x_i^{(2)}}{, \quad \forall i}{\nonumber}
		\addConstraint{x_i^{(3)}}{\in \mathbb{Z}^+}{, \quad \forall i}{\nonumber}
	\end{mini!}
	
	% \paragraph{Model B: Maximize Coverage}
	% \begin{maxi!}|l|
	% 	{x_1^{(3)}, \ldots, x_m^{(3)}}{\sum_{i=1}^{m} x_i^{(3)}}{}{}\tag*{}
	% 	\addConstraint{\sum_{i=1}^{m} x_i^{(3)}}{\leq 2n}{}{\nonumber}
	% 	\addConstraint{(1-\rho)\bar{x}^{(3)}}{\leq x_i^{(3)} \leq (1+\rho)\bar{x}^{(3)}}{, \quad \forall i}{\nonumber}
	% 	\addConstraint{x_i^{(3)}}{\leq x_i^{(2)}}{, \quad \forall i}{\nonumber}
	% 	\addConstraint{x_i^{(3)}}{\in \mathbb{Z}_+}{, \quad \forall i}{\nonumber}
	% \end{maxi!}
	
	The key difference from Round 2 is the sequential constraint $x_i^{(3)} \leq x_i^{(2)}$, ensuring Round 3 quotas do not exceed Round 2 quotas, and the tighter coverage bounds reflecting that PNMs rank up to 2 chapters in Round 3.
	
	\subsection{Step 4: Final Matching and Capacity Determination}
	
	After Round 3, the final matching phase begins. PNMs sign a Membership Recruitment Acceptance Binding Agreement (MRABA), committing to accept any bid they receive from a chapter they ranked. Each PNM then ranks up to 2 chapters from those they visited in Round 3.	Simultaneously, each chapter submits a ranked preference list of all PNMs they invited in Round 3.\\
	
	Let $\succ_p$ denote PNM $p$'s preference ordering over chapters (ranking up to 2), and $\succ_{c_i}$ denote chapter $c_i$'s preference ordering over PNMs. Some PNMs employ a suicide bid strategy, ranking only their top-choice chapter rather than ranking 2 chapters. If their top choice does not select them, they remain unmatched. This strategy reflects strong preference intensity but carries significant risk. Our proposed system, which integrates ML-based compatibility prediction, invitation quota optimization, and stable matching algorithms, aims to maximize overall match rates, thereby reducing the number of unmatched PNMs including those employing suicide bids.\\

	We consider the graph algorithm, the Deferred Acceptance (DA) algorithm \cite{gale1962college}. Below, we address it. 
    
    %and the Hungarian algorithm (Maximum Weight Matching) \cite{kuhn1955hungarian}. In the following we mention both of the separately. 
	
	\subsubsection{Deferred Acceptance Algorithm}
	
	 The DA algorithm guarantees a stable matching $\mu: \mathcal{P} \rightarrow \mathcal{C} \cup \{\emptyset\}$, where $\mu(p) = c$ indicates PNM $p$ is matched to chapter $c$, and $\mu(p) = \emptyset$ indicates $p$ is unmatched. Stability means there exists no blocking pair $(p, c)$ where both $p$ and $c$ would prefer each other over their current matches. The DA algorithm operates as follows.
	
	\begin{algorithm}[H]
		\caption{Deferred Acceptance (DA) Algorithm}
		\begin{algorithmic}[1]
			\STATE Initialize all PNMs and chapters as unmatched
			\WHILE{$\exists$ unmatched PNM $p$ who has not proposed to all ranked chapters}
			\STATE PNM $p$ proposes to their most-preferred chapter $c$ not yet proposed to
			\IF{Chapter $c$ is unmatched}
			\STATE $c$ tentatively accepts $p$
			\ELSIF{$c$ prefers $p$ over current tentative match $p'$}
			\STATE $c$ rejects $p'$ and tentatively accepts $p$
			\STATE $p'$ becomes unmatched
			\ELSE
			\STATE $c$ rejects $p$
			\ENDIF
			\ENDWHILE
			\STATE All tentative matches become final
		\end{algorithmic}
	\end{algorithm}

	\subsubsection{Capacity Determination}
	
	A key feature of this recruitment system is that chapter capacities are not predetermined. Instead, capacities emerge from the matching process itself. Regardless of which matching algorithm is used (DA or Hungarian), the realized capacity of chapter $c_i$ is
	\begin{equation*}
		\text{cap}_i = |\{p \in \mathcal{P} : \mu(p) = c_i\}|
	\end{equation*}
	
	Paula's goal is to ensure that nearly all PNMs receive bids and are matched, which effectively determines the chapter capacities needed. We verify that realized capacities satisfy the fairness constraint
	\begin{equation*}
		(1 - \rho)\bar{\text{cap}} \leq \text{cap}_i \leq (1 + \rho)\bar{\text{cap}}, \quad \forall i
	\end{equation*}
	where $\bar{\text{cap}} = \frac{1}{M}\sum_{i=1}^{M}\text{cap}_i$.
	
	If the matching produced by either algorithm violates fairness constraints, we implement a capacity-constrained variant where each chapter has upper and lower capacity bounds set according to $(1-\rho)\bar{\text{cap}}$ and $(1+\rho)\bar{\text{cap}}$, and the algorithm is run iteratively, adjusting bounds until all PNMs are matched within fairness constraints.

\section{Experimental Results}
\label{sec:results}

% This section presents the empirical evaluation of our ML-guided quota optimization framework applied to sorority recruitment data from a small Midwestern university. Our analysis proceeds in four steps, each corresponding to a component of the
% framework described in Section~4. Due to data availability constraints, Steps~1 and~2 (compatibility prediction and quota optimization) are evaluated using 2021--2023 as training data and 2024 as the test year, a period during which four
% chapters (A, B, C, and D) participated in recruitment. Step~4 (final matching via Deferred Acceptance) is evaluated on the 2025 recruitment cycle, in which Chapter~D had become inactive and all historical records were restructured to reflect a
% three-chapter system. We discuss the implications of this split evaluation in Section~\ref{sec:limitations}. The 2024 cohort consisted of 82 registered PNMs, and the 2025 cohort consisted of 56 PNMs who completed the full recruitment process
% through Preference Night.

This section presents the empirical evaluation of our ML-guided quota optimization framework applied to sorority recruitment data from a small 
Midwestern university. Our analysis proceeds in four steps, each corresponding to a component of the framework described in Section~4. We begin by describing the data and evaluation strategy, then present results for each component 
in turn. The 2024 cohort consisted of 82 registered PNMs, and the 2025 cohort consisted of 56 PNMs who completed the full recruitment process 
through Preference Night.

\subsection{Data and Evaluation Strategy}
\label{sec:eval_strategy}

A key structural feature of our empirical evaluation is that it spans two 
distinct recruitment cohorts. Steps~1 and~2 of the framework (ML-based 
compatibility prediction and invitation quota optimization) are evaluated 
using 2021--2023 as training data and 2024 as the test year, a period during 
which four chapters (A, B, C, and D) participated in recruitment. Step~4 
(final matching via Deferred Acceptance) is evaluated on the 2025 recruitment 
cycle, the first year in our dataset with a fully stable three-chapter structure 
following Chapter~D's withdrawal.\\

This split arises from two factors. First, chapter participation changed 
between cohorts, as Chapter~D withdrew after the 2024 cycle. Second, complete 
Round~2 preference data were unavailable for the 2024 cohort, which precluded 
a fully integrated end-to-end evaluation on a single cohort. We treat the two 
evaluations as complementary. The 2024 results validate the optimization 
components in isolation, while the 2025 results validate the matching component 
against a complete and consistent preference dataset. Together they provide 
component-level empirical support for the framework under realistic data 
constraints.\\

A fully integrated evaluation applying ML-optimized quotas and evaluating 
the resulting matches on the same cohort remains an important direction 
for future work. Such an evaluation would require a recruitment cycle for 
which complete registration data, Round~2 preference rankings, Round~3 bid 
lists, and final MRABA preferences are all available within a consistent 
chapter structure.

% -------------------------------------------------------
\subsection{ML-Based Compatibility Prediction}
\label{sec:results_ml}
% -------------------------------------------------------

We trained a Random Forest binary classifier on historical PNM--chapter pair data from 2021--2023 to predict the probability that a given PNM would be matched to a given chapter. The training set was constructed by enumerating all PNM--chapter
pairs for each recruitment year and labeling each pair as a match ($y = 1$) or non-match ($y = 0$) based on final recruitment outcomes. This procedure yielded 1,128 training pairs in total: 282 matched pairs and 846 non-matched pairs,
producing a 3:1 class imbalance with non-matches comprising 75\% of observations. Sixteen PNMs had missing high school GPA values; these were imputed using the training set median of 3.90. The two derived features dependent on GPA (GPA differential and the GPA-above-chapter indicator) were imputed accordingly. Model selection used group $k$-fold cross-validation ($k = 5$) stratified by PNM identity to prevent data leakage across pairs derived from the same individual.\\

The Random Forest model was trained on 13 features: 10 numerical variables and 3 categorical variables (major, residence hall, and chapter name). Hyperparameter tuning over $\text{mtry} \in \{3, 5, 7\}$ selected $\text{mtry} = 7$ as optimal, yielding a cross-validated ROC-AUC of 0.5822, sensitivity of 0.9973, and specificity of 0.000.\\

The near-perfect sensitivity (0.997) combined with specificity of zero indicates that the model collapses to predicting every pair as a non-match at its default threshold, a well-known consequence of severe class imbalance without resampling
or threshold adjustment \cite{he2009learning}. Nevertheless, the ROC-AUC of 0.5822 confirms that the model's predicted probabilities rank true matches above non-matches more often than chance, which is precisely what our framework requires: we use $S_{pc} \in [0,1]$ as relative compatibility signals for quota optimization, not as binary classifiers. The modest AUC reflects both the inherent difficulty of predicting social 
compatibility from pre-recruitment registration data and the severe data limitations of this setting. With only 282 positive training instances drawn from a single small institution over three years, meaningful discrimination is unlikely regardless of model choice.\\

\begin{figure}[H]
	\centering
	\includegraphics[width=0.5\textwidth]{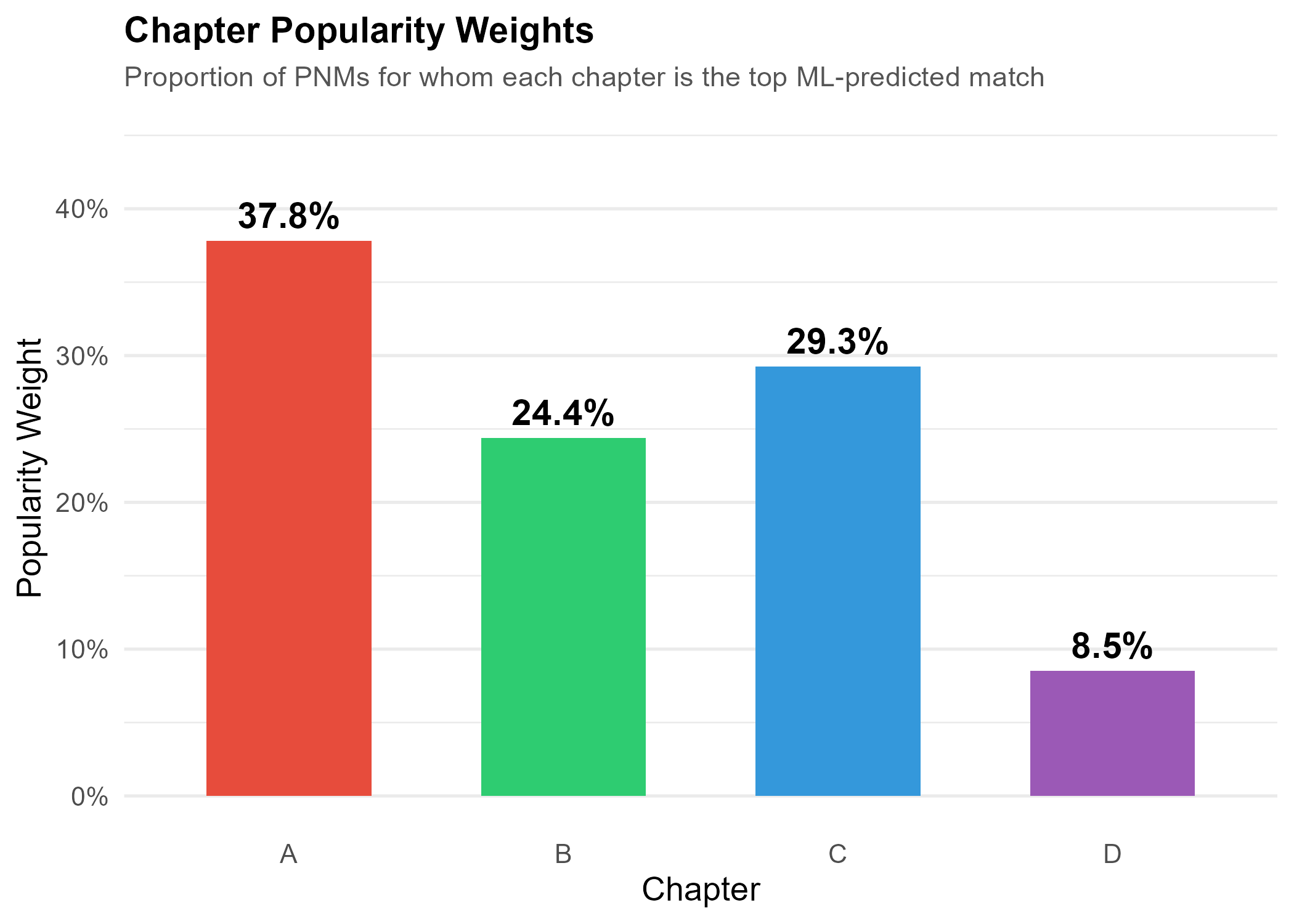}
	\caption{Chapter popularity weights $w^{(2)}_i$ for the 2024 PNM cohort ($n = 82$). Each bar represents the proportion of PNMs for whom that chapter received the highest predicted compatibility score. Chapter~A is the plurality top-predicted match (37.8\%), followed by Chapter~C (29.3\%), Chapter~B (24.4\%), and Chapter~D (8.5\%).}
	\label{fig:popularity_weights}
\end{figure}

\begin{figure}[H]
	\centering
	\includegraphics[width=0.7\textwidth]{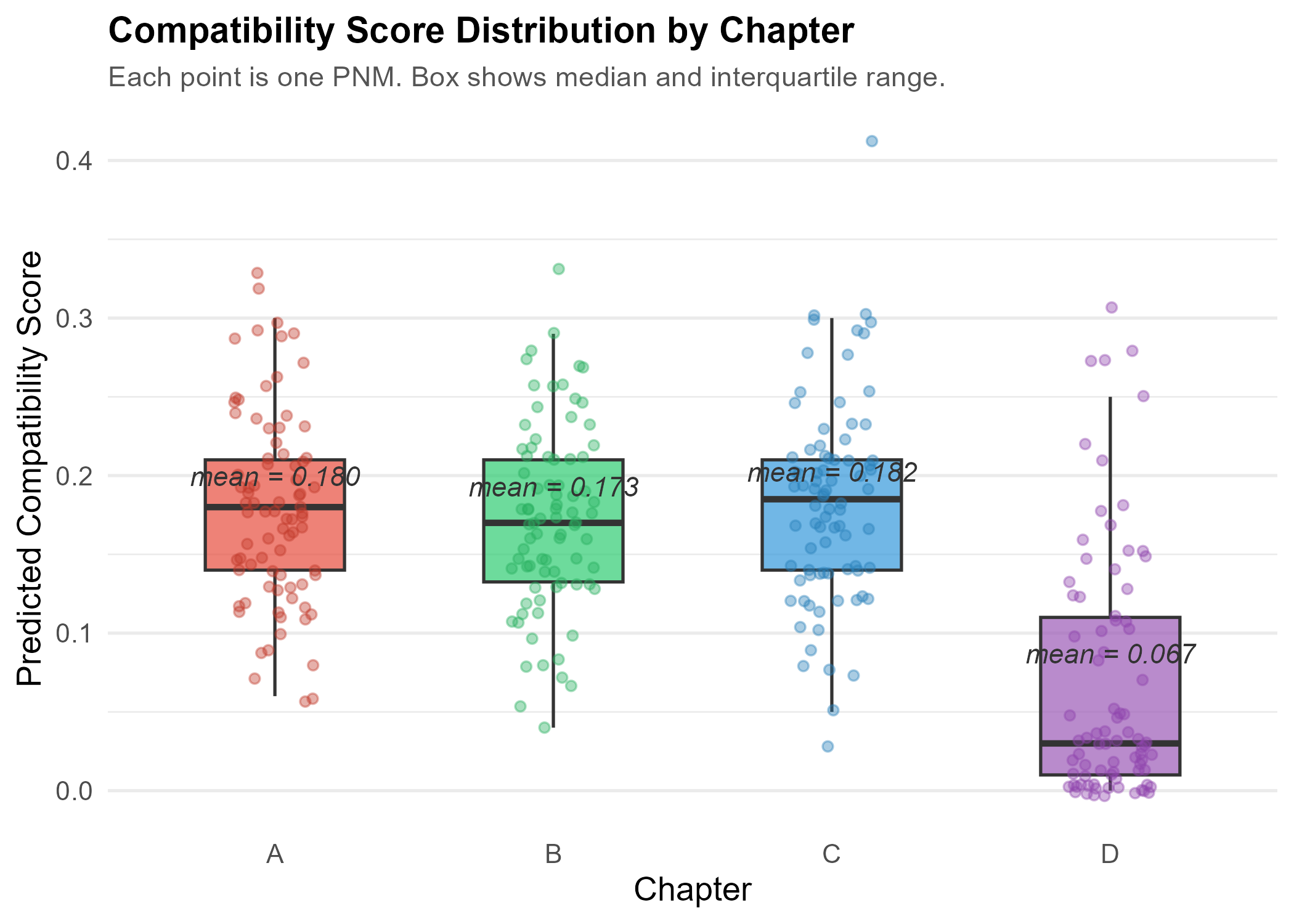}
	\caption{Distribution of predicted compatibility scores across all 82 PNMs for each chapter. Each point represents one PNM--chapter pair. Chapters~A, B, and C exhibit similar mean scores of 0.180, 0.173, and 0.182 respectively,
		while Chapter~D shows a substantially lower mean of 0.067 with scores concentrated near zero, reflecting its declining participation leading up to the 2024 cycle.}
	\label{fig:score_distribution}
\end{figure}

\begin{figure}[H]
	\centering
	\includegraphics[width=0.75\textwidth]{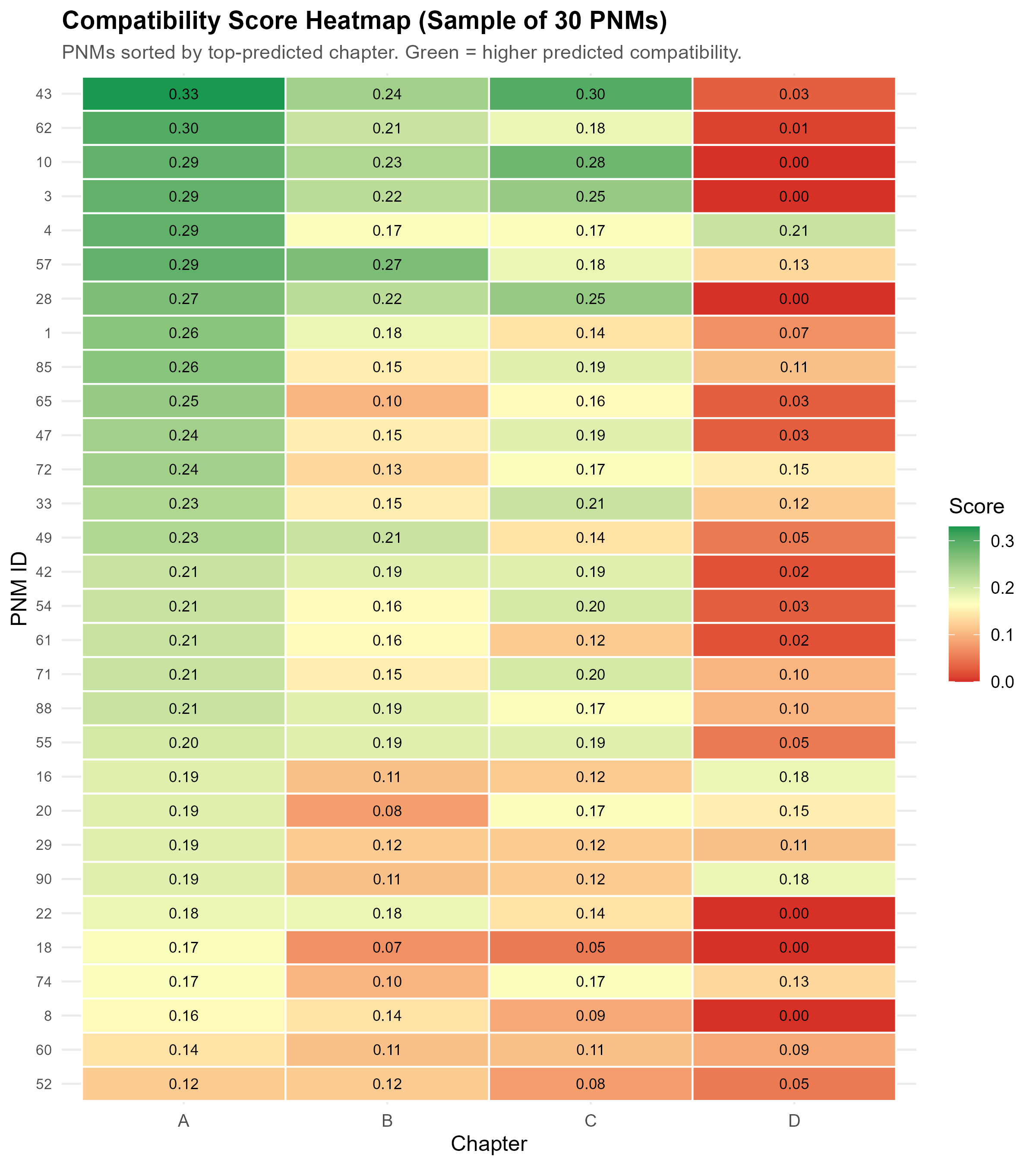}
	\caption{Compatibility score heatmap for a proportionally sampled subset of 30 PNMs from the 2024 cohort ($n = 82$), grouped and sorted by top-predicted chapter. Each cell reports the predicted probability $S_{pc}$ that PNM $p$
		is compatible with chapter $c$. Green shading indicates higher predicted compatibility; red indicates lower. Chapter~D's near-zero scores (deep red column) are visible across nearly all groups. The highest single score in the sample belongs to PNM~79 (Chapter~C: 0.41).}
	\label{fig:heatmap}
\end{figure}

The resulting compatibility matrix $S \in \mathbb{R}^{82 \times 4}$ for the 2024 cohort showed mean predicted scores of 0.180, 0.173, and 0.182 for Chapters~A, B, and C respectively, while Chapter~D exhibited a substantially lower mean of
0.067 with scores concentrated near zero. This sharp divergence in Chapter~D's predicted compatibility further corroborates its approaching inactivity in 2024. Figure~\ref{fig:popularity_weights} displays the ML-derived popularity weights $w^{(2)}_i$ for each chapter. Figure~\ref{fig:score_distribution} shows the full
distribution of predicted scores across all PNMs for each chapter, and Figure~\ref{fig:heatmap} presents a proportionally sampled subset of 30 PNMs from the compatibility matrix $S$, grouped and sorted by top-predicted chapter.

% --- Figure 1: Popularity Weights ---

% --- Figure 2: Score Distribution ---

% --- Figure 3: Heatmap ---

% -------------------------------------------------------
\subsection{Round 2 Invitation Quota Optimization}
\label{sec:results_round2}
% -------------------------------------------------------

Using the 2024 compatibility matrix, we computed ML-derived popularity weights $w^{(2)}_i$ for each chapter and solved the Round~2 integer linear program described in Section~4.3. The fairness parameter $\rho = 0.2676$ was derived from historical
capacity variation across 2021--2023 (Chapter~A: 41 new members, B: 37, C: 38, D: 26), permitting chapter quotas to deviate by at most $26.8\%$ from their mean.\\

\begin{figure}[H]
	\centering
	\includegraphics[width=0.65\textwidth]{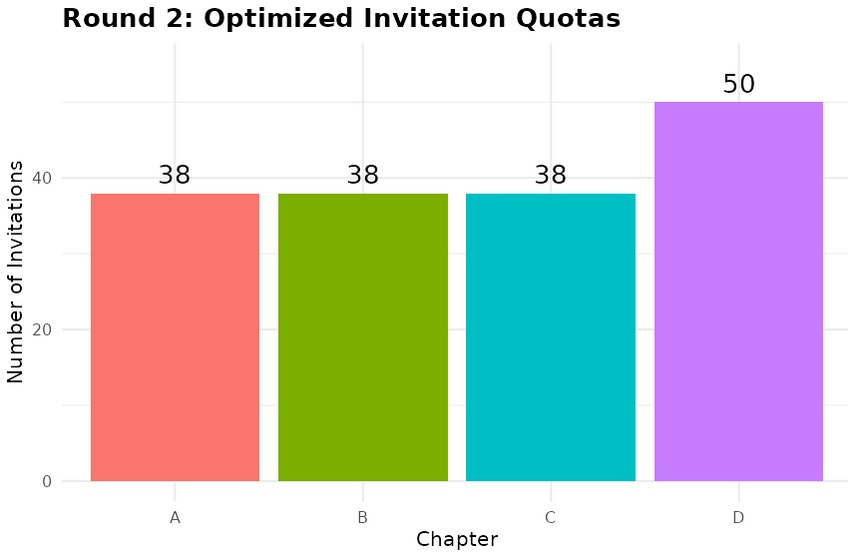}
	\caption{Optimized Round~2 invitation quotas by chapter under the efficiency-focused model. Chapters~A, B, and C each receive a quota of 38, while Chapter~D receives a higher quota of 50 due to its lower popularity
		weight ($w^{(2)}_D = 0.085$).}
	\label{fig:round2_quotas}
\end{figure}

Table~\ref{tab:round2_results} summarizes the popularity weights, optimized quotas, and actual 2024 quotas for comparison. The optimized Round~2 quotas were 38 for
Chapters~A, B, and C, and 50 for Chapter~D, totaling 164 invitations across 82 PNMs, an average of 2.0 invitations per PNM. The higher quota assigned to Chapter~D reflects the model's compensatory logic: because Chapter~D had the lowest
popularity weight ($w^{(2)}_D = 0.085$), the efficiency-focused objective directed additional invitation capacity toward it to ensure sufficient PNM coverage despite weaker predicted appeal. Figure~\ref{fig:round2_quotas} displays the optimized
quotas by chapter.

\begin{table}[H]
	\centering
	\caption{Round 2: Popularity Weights and Invitation Quotas (2024)}
	\label{tab:round2_results}
	\begin{tabular}{lcccc}
		\hline
		\textbf{Chapter} & \textbf{Popularity Weight} & \textbf{Optimized Quota} &
		\textbf{Actual Quota} & \textbf{Difference} \\
		\hline
		A & 0.3780 & 38 & 38 & $~~$0 \\
		B & 0.2439 & 38 & 42 & $+$4 \\
		C & 0.2927 & 38 & 37 & $-$1 \\
		D & 0.0854 & 50 &  0 & $-$50 \\
		\hline
		\textbf{Total} & -- & 164 & 117 & -- \\
		\hline
	\end{tabular}
\end{table}

% --- Figure 4: Round 2 Optimized Quotas ---

\begin{figure}[H]
	\centering
	\includegraphics[width=0.65\textwidth]{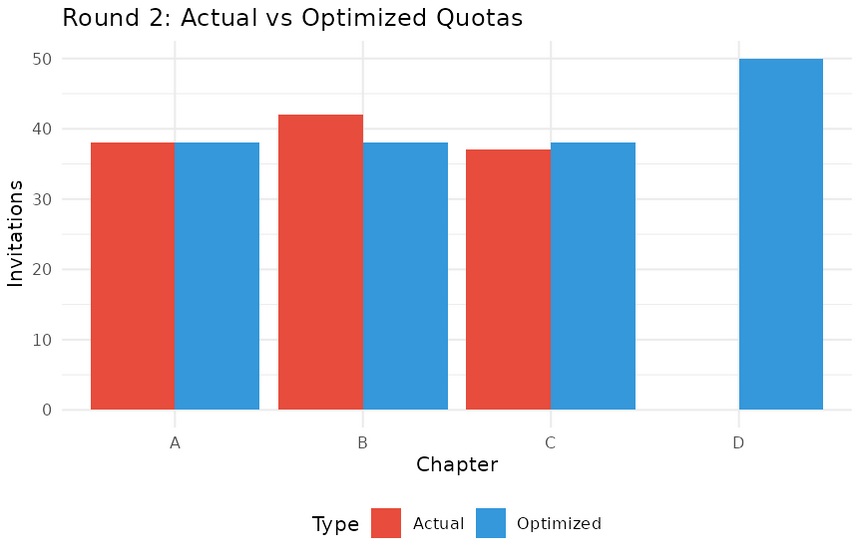}
	\caption{Round~2 optimized versus actual invitation quotas for the 2024 recruitment cycle. Chapter~A matches exactly (38 vs.\ 38), Chapter~B differs by $+$4 (42 actual vs.\ 38 optimized), and Chapter~C differs by $-$1 (37 actual vs.\ 38 optimized). The absence of an actual bar for
		Chapter~D reflects its inactivity in 2024.}
	\label{fig:round2_comparison}
\end{figure}

Comparing the optimized quotas against actual 2024 coordinator decisions, the optimizer's recommendations for the three active chapters (A, B, C) differed from actual quotas by 0, $+$4, and $-$1 invitations respectively, as shown in
Figure~\ref{fig:round2_comparison}.

% --- Figure 5: Round 2 Actual vs Optimized ---

The large discrepancy for Chapter~D (optimized: 50, actual: 0) reflects the framework's inability, using training data from 2021--2023, to anticipate Chapter~D's withdrawal from the 2024 recruitment cycle. This is a known limitation that arises when the chapter participation set changes between training and test
years, and is addressed in future work by incorporating a chapter activity flag as a preprocessing step.

% -------------------------------------------------------
\subsection{Round 3 Invitation Quota Optimization}
\label{sec:results_round3}
% -------------------------------------------------------

\begin{figure}[H]
	\centering
	\includegraphics[width=0.65\textwidth]{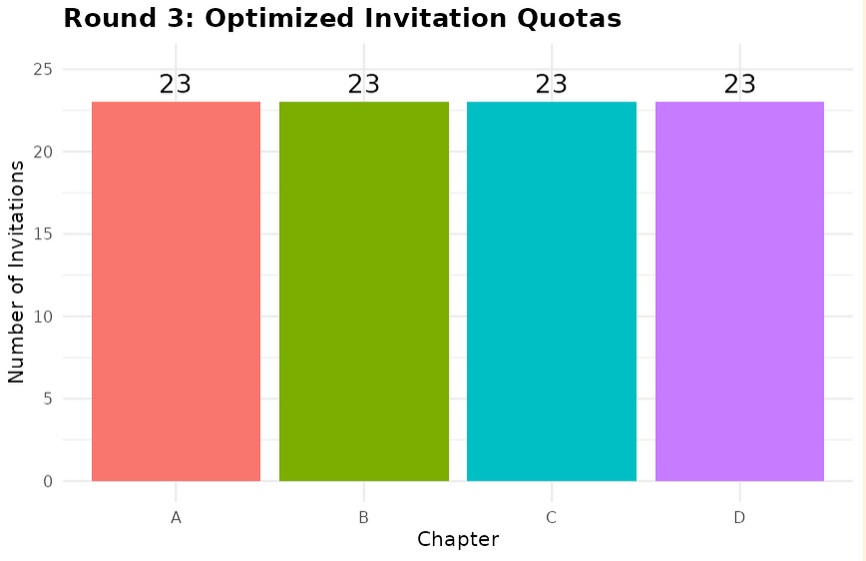}
	\caption{Optimized Round~3 invitation quotas by chapter. All four chapters receive equal quotas of 23. Reductions from Round~2 are 15 invitations each for Chapters~A, B, and C, and 27 for Chapter~D.}
	\label{fig:round3_quotas}
\end{figure}

For Round~3, reciprocal rates $r_i$ were computed from simulated Round~2 PNM rankings derived from compatibility scores, since actual Round~2 ranking data were unavailable for the 2024 cohort. This simulation assigns multiple chapters
as top preferences for most PNMs, producing reciprocal rates exceeding 1.0 for popular chapters (A: 2.079, B: 1.895, C: 2.026) --- a known artifact we acknowledge as an implementation limitation. Despite this, the Round~3 optimization produced well-defined quotas of 23 for each chapter, totaling 92
invitations (average 1.12 per PNM), with reductions of 15 from Round~2 for Chapters~A, B, and C, and 27 for Chapter~D. The uniform quota reflects the interplay between the fairness constraint ($\rho = 0.2676$) and coverage bounds.
Figure~\ref{fig:round3_quotas} and Table~\ref{tab:round3_results} display these
results.

\begin{table}[H]
	\centering
	\caption{Round 3: Reciprocal Rates and Invitation Quotas (2024)}
	\label{tab:round3_results}
	\resizebox{\textwidth}{!}{%
		\begin{tabular}{lccccc}
			\hline
			\textbf{Chapter} & \textbf{Reciprocal Rate} & \textbf{Round 2 Quota} &
			\textbf{Round 3 Quota} & \textbf{Actual Quota} & \textbf{Difference} \\
			\hline
			A & 2.079 & 38 & 23 & 23 & $~~$0 \\
			B & 1.895 & 38 & 23 & 26 & $+$3 \\
			C & 2.026 & 38 & 23 & 21 & $-$2 \\
			D & 0.360 & 50 & 23 &  0 & $-$23 \\
			\hline
			\textbf{Total} & -- & 164 & 92 & 70 & -- \\
			\hline
		\end{tabular}%
	}
\end{table}

% --- Figure 6: Round 3 Optimized Quotas ---

\begin{figure}[H]
	\centering
	\includegraphics[width=0.65\textwidth]{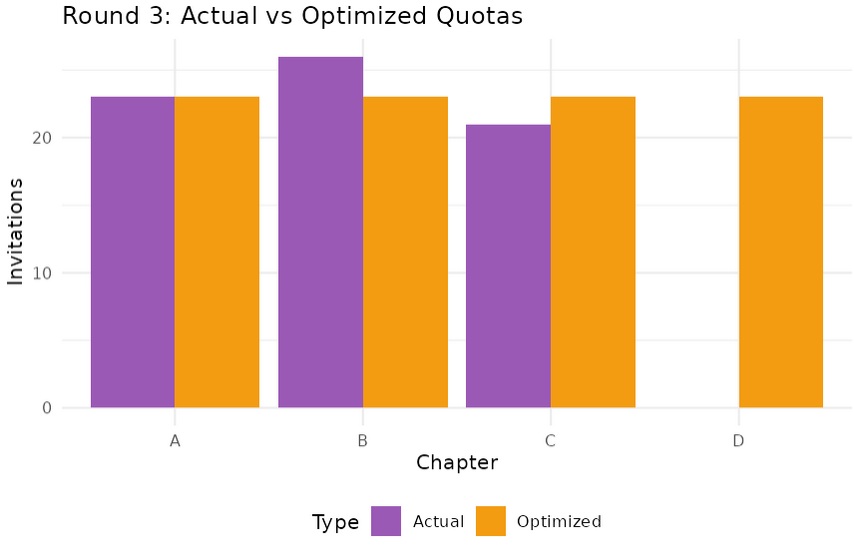}
	\caption{Round~3 optimized versus actual invitation quotas for the 2024 recruitment cycle. Chapter~A matches exactly (23 vs.\ 23), Chapter~B differs by $+$3 (26 actual vs.\ 23 optimized), and Chapter~C differs by $-$2 (21 actual vs.\ 23 optimized). The absence of an actual bar for
		Chapter~D reflects its inactivity in the 2024 cycle.}
	\label{fig:round3_comparison}
\end{figure}

Among the three active chapters, the optimizer matched Chapter~A exactly (23 vs. 23), and differed by $+$3 and $-$2 for Chapters~B and C respectively, as shown in Figure~\ref{fig:round3_comparison}. The discrepancy for Chapter~D is again attributable entirely to its inactivity in 2024.

% --- Figure 7: Round 3 Actual vs Optimized ---

% -------------------------------------------------------
\subsection{Final Matching Outcomes}
\label{sec:results_matching}
% -------------------------------------------------------

We evaluate the Deferred Acceptance algorithm on the 2025 recruitment cycle, the most recent year for which complete preference and outcome data are available across
a consistent three-chapter system. As noted above, Chapter~D's records were removed from the dataset by 2025, making this the first recruitment year in our dataset with a fully stable three-chapter structure. The 2025 cohort consisted of 56 PNMs
who completed the full recruitment process through Preference Night and submitted MRABA preferences across Chapters~A, B, and C.\\

\begin{figure}[H]
	\centering
	\includegraphics[width=0.65\textwidth]{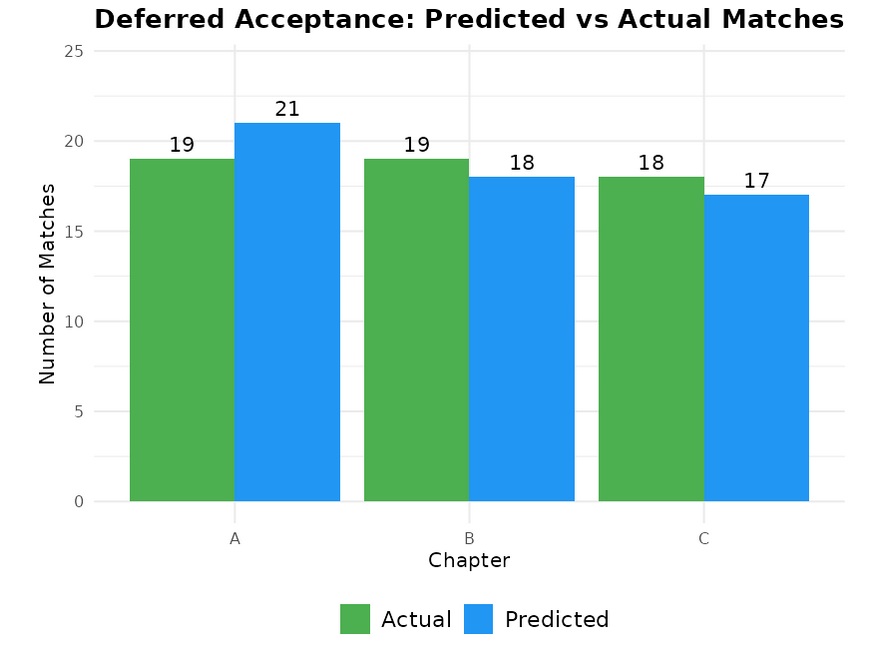}
	\caption{Deferred Acceptance algorithm: predicted versus actual new member counts by chapter for the 2025 recruitment cycle ($n = 56$). The maximum chapter-level deviation is 2 members (Chapter~A), and predicted totals
		match actual totals exactly (56 vs.\ 56). Green bars show actual outcomes; blue bars show DA predictions.}
	\label{fig:matching_comparison}
\end{figure}

The DA algorithm achieved a 100\% match rate, with all 56 participating PNMs successfully matched to a chapter, with zero unmatched outcomes. This result mirrors the actual 2025 recruitment outcome, in which all 56 PNMs who reached
Preference Night also received bids. Table~\ref{tab:matching_results} reports
predicted and actual match counts by chapter, and Figure~\ref{fig:matching_comparison}
presents a visual comparison.

\begin{table}[H]
	\centering
	\caption{Final Matching Outcomes: Predicted vs.\ Actual (2025)}
	\label{tab:matching_results}
	\begin{tabular}{lccc}
		\hline
		\textbf{Chapter} & \textbf{Predicted (DA)} & \textbf{Actual} &
		\textbf{Difference} \\
		\hline
		A & 21 & 19 & $-$2 \\
		B & 18 & 19 & $+$1 \\
		C & 17 & 18 & $+$1 \\
		\hline
		\textbf{Total} & \textbf{56} & \textbf{56} & \textbf{0} \\
		\hline
	\end{tabular}
\end{table}

% --- Figure 8: DA Predicted vs Actual Matches ---

The maximum chapter-level deviation was 2 new members (Chapter~A), and predicted totals matched actual totals exactly. At the individual level, the DA algorithm's
assignment agreed with actual recruitment outcomes for 54 of 56 PNMs, a 96.4\% agreement rate.

% % -------------------------------------------------------
% \subsection{Data Evaluation Strategy and Limitations}
% \label{sec:limitations}
% % -------------------------------------------------------

%  Steps~1--2 (ML prediction and quota optimization) are evaluated on 2024 data, while Step~4 (final matching) is evaluated on 2025 data. This split arises because
%  Chapter~D became inactive after 2024 and the 2025 dataset was restructured to a three-chapter system, making it the cleanest setting for evaluating the matching algorithm. As a result, a fully integrated end-to-end evaluation, applying
%  ML-optimized quotas and evaluating the resulting matches on the same cohort, was not possible with the available data, and the 2025 matching uses actual 2025 invitation and preference data rather than optimizer-generated quotas. This remains
%  an important direction for future work. Despite this limitation, the individual components of the framework perform credibly in isolation. The optimizer produces quota recommendations that closely align with actual coordinator decisions for active chapters in both rounds. The DA algorithm replicates actual match outcomes with a 96.4\% individual-level agreement rate and a 100\% match rate. Together, these results provide meaningful empirical support for the framework's core design, while honestly acknowledging the constraints of the available data.

\subsection{Limitations}
\label{sec:limitations}

As described in Section~\ref{sec:eval_strategy}, the data constraints of this setting precluded a fully integrated end-to-end evaluation on a single cohort. Despite this, the individual components of the framework perform credibly in isolation. The optimizer produces quota recommendations that 
closely align with actual coordinator decisions for active chapters in both rounds, and the Deferred Acceptance algorithm replicates actual match outcomes with a 96.4\% individual-level agreement rate and a 100\% match rate. Together, 
these results provide meaningful empirical support for the framework's core design while honestly acknowledging the constraints of the available data.

\section{Conclusion}

This study presents an integrated framework combining ML-based compatibility prediction with integer programming for invitation quota optimization in small 
two-sided matching markets, applied to five years of sorority recruitment data from a small Midwestern university. An interactive web application enables 
recruitment coordinators to implement these methods in practice.\footnote{Available at \url{https://rstudio.evansville.edu/Sorority_matching/}} Empirical evaluation demonstrates that the framework performs credibly in isolation across its components. The Random Forest compatibility model achieves a cross-validated ROC-AUC of 0.5822, providing a structured signal for quota allocation despite the inherent difficulty of predicting social compatibility from pre-recruitment data. The optimized invitation quotas closely align with actual coordinator decisions for active chapters in both rounds. The Deferred
Acceptance algorithm replicates actual 2025 recruitment outcomes with a 96.4\% individual-level agreement rate and a 100\% match rate across 56 PNMs. As described in Section~\ref{sec:limitations}, component-level evaluation under realistic data constraints provides meaningful support for the 
framework's design. Future work should pursue a fully integrated evaluation within a single consistent recruitment cycle, and explore dynamic learning approaches, improved compatibility features collected during recruitment, 
and generalization to other constrained matching environments including school choice, residency matching, and thin labor markets.

\end{document}